\documentclass[a4paper,12pt,preprint]{revtex4}
\usepackage{epsf}
\usepackage{graphicx}
\usepackage{amssymb,amsmath}

\DeclareMathAlphabet{\mathsfsl}{OT1}{cmss}{m}{sl}

\begin{document}

\renewcommand{\tensor}[1]{\mathsfsl{#1}}
\renewcommand{\vec}[1]{\mbox{\boldmath$#1$}}

\title{Bounds on Diatomic Molecules in a Relativistic Model}

\author{Natalie Gilka}
\affiliation{Department of Mathematical Sciences,
  University of Copenhagen, Universitetsparken 5, 2100 Copenhagen
  {\O}, Denmark}

\begin{abstract}
  We consider diatomic systems in which the kinetic energy of the
  electrons is treated in a simple relativistic model. The
  Born--Oppenheimer approximation is assumed. We investigate questions
  of stability, deducing bounds on the number $N$ of electrons, the
  binding energy $\Delta E_b$ and the equilibrium bond distance
  $R_0$. We use a known localization argument adopted to the present
  relativistic setting, with particular consideration of the critical
  point of stability, as well as the recently proved relativistic
  Scott correction.
\end{abstract}

\maketitle

\setlength{\arraycolsep}{0.15em}

\section{Introduction}

In this paper we study the stability of diatomic systems in the
Born--Oppenheimer formulation. The kinetic energy of the electrons is
treated in a model which accounts for relativistic effects. We employ
a known localization argument applied by Solovej\cite{Sol90} to prove
a lower bound on the number $N$ of electrons. This involves
establishing an estimate of the equilibrium bond distance $R_0$ as a
function of $N$ which requires particular consideration in the
relativistic case due to the critical point of stability for large
values of the nuclear charges. Bounds on the binding energy $\Delta
E_b$ and on $R_0$ are deduced using the recently proved relativistic
Scott correction\cite{Sol10}.\\

We introduce positive numbers $Z_1, Z_2$, $\vec{Z} = (Z_1, Z_2)$,
which correspond in a physical picture to nuclear charges of a
diatomic molecule, and denote the nuclear coordinates as $\vec{R}_1$,
$\vec{R}_2 \in \mathbb{R}^3$. We consider furthermore a set of $N$
electrons and introduce the electron coordinates $x_i \in
\mathbb{R}^3$, $i=1, \: \ldots, \: N$. The coordinate origin is
placed at $\frac{1}{2}(\vec{R}_1+\vec{R}_2)$. Let
$\vec{R}_1 = \vec{R}/2$, $\vec{R}_2 = -\vec{R}/2$, where
$\vec{R} = \vec{R}_1 - \vec{R}_2$, furthermore $R =
|\vec{R}_1-\vec{R}_2|$.
We are working in units where $m = e = \hbar = 1$ ($m$: electron mass;
$e$: electron charge; $\hbar$: Planck's constant). In this unit
system, the distance is measured in Bohr radii $a_0 = \hbar^2 /m
e^2$. The Hamiltonian employed in the present context is then
\begin{eqnarray}
\label{H1}
H(N,\vec{Z},\vec{R};\alpha) & = &
\sum_{i=1}^N \left( \sqrt{-\alpha^{-2} \, \Delta_i + \alpha^{-4}} - \alpha^{-2} \right)
- \sum_{i=1}^N \left( \frac{Z_1}{\left| x_i - \vec{R}_1 \right|}
+ \frac{Z_2}{\left| x_i - \vec{R}_2 \right|} \right) \nonumber \\
& & + \sum_{i<j} \frac{1}{\left| x_i - x_j \right|}
+ \frac{Z_1 Z_2}{R} \nonumber \\
& = & H^{\alpha} + H_{en} + H_{ee} + H_{nn}.
\end{eqnarray}
$\alpha$ denotes the dimensionless fine-structure constant. It can be
viewed as a quantifier of the relativistic correction and is in our
unit system defined as $\alpha = c^{-1}$, where $c$ denotes the speed
of light. In the nonrelativistic limit, we have $\alpha \rightarrow
0$. We assume the Born--Oppenheimer approximation, i.e., fixed nuclei
positions $\vec{R}_1$, $\vec{R}_2$ corresponding to infinite nuclear
masses $M_1 = M_2 = \infty$, and obtain therefore a parametrical
dependence of the Hamiltonian on $\vec{R}$. \\

The terms $H_{en}$, $H_{ee}$, $H_{nn}$ denote two-particle operators
describing the classical Coulomb interaction between electrons $(e)$
and nuclei $(n)$. The term $ H^{\alpha}$ in (\ref{H1}) is a
one-particle operator considering the kinetic energy of electrons of
mass $m$, with $\Delta_i$ referring to the Laplacian of the $i$-th
electron. This form of the operator represents the simplest model that
attempts to include relativistic effects\cite{Wed75,Herb77,Lieb88}. It
does not give accurate numerical agreement with experimental
observations, however, it does provide a qualitatively sensible
description. For $\alpha = 0$ we retrieve the nonrelativistic limit
of the kinetic energy of the $i$-th electron of
$-\frac{1}{2}\Delta$. The simple model chosen here has in common with
all other relativistic approaches that it predicts instability of
large atoms and molecules. The relevant parameter in this context is
the product of the atomic numbers $Z_k$, $k = 1,2$, and the
fine-structure constant $\alpha$. The correct critical value of $Z_k
\alpha$ as derived from the study of the Dirac equation is expected to
assume $Z_k \alpha = 1$. If the interest lies in the consideration of
the limiting behaviour for $Z_k \rightarrow \infty$, one is therefore
required to bound $Z_k \alpha$ by imposing $\alpha \rightarrow 0$, as
was done in \cite{Sol10}. While this is from a physical perspective
not sensible as the fine-structure constant exhibits an experimentally
established value of approximately $1/137$, it permits though a
mathematical consideration of the asymptotics. The above discussion
referred to the case of a diatomic system; the same is true for
more complicated models.\\

Concerning the abovementioned relativistic instability
quantitatively, we recognize that our simple relativistic model
exhibits a known deficiency in that its point of instability is in
contradiction with the expected value of $Z_k \alpha = 1$. In our model,
the ground state energy $E\left( N, \vec{Z}, R; \alpha \right)$ is
finite if $\underset{k}{\mbox{max}} \, \{Z_k \alpha\} \le 2/\pi$, but $E( N,
\vec{Z}, R; \alpha ) = - \infty$ if $\underset{k}{\mbox{max}} \, \{Z_k
  \alpha\} > 2/\pi$ (see \cite{Wed74,Con84,Feff86}, as well as
\cite{Herb77,Lieb88,Sol10} for further discussion). This imposes that
the atomic number has to be smaller than or equal to $2/\pi \alpha
\approx 87$, thereby contradicting the observation of atoms with
atomic numbers larger than 87 to be stable.  It is therefore clear
that while our model is qualitatively reliable,
we cannot expect good quantitative agreement. \\

The operator $H(N,\vec{Z},\vec{R};\alpha)$ acts on the
fermionic space $\bigwedge\limits_{i=1}^N L^2 \left( \mathbb{R}^3;
 \mathbb{C}^2 \right)$. The energy
$E(N, \vec{Z}, R; \alpha)$ as a function of $R$ is given as
\begin{eqnarray*}
E\left( N, \vec{Z}, R; \alpha \right) = \inf \: \mbox{spec} \: H(N,
\vec{Z}, \vec{R}; \alpha).  
\end{eqnarray*}
We introduce the energy minimum (or rather infimum) of the molecule
\begin{eqnarray*}
E\left( N, \vec{Z}; \alpha \right) = \inf_R E(N, \vec{Z}, R; \alpha).
\end{eqnarray*}
That $E\left( N, \vec{Z}; \alpha \right) > - \infty$, i.e., the
stability of relativistic molecules, was first proved by Daubechies
and Lieb\cite{Daub83}. A more general proof is due to
Conlon\cite{Con84} and Fefferman and de~la~Lave\cite{Feff86}.\\

We say that the molecule has a stable Born--Oppenheimer ground state if
the following two requirements are satisfied. 
\renewcommand{\labelenumi}{(\arabic{enumi}).}
\begin{enumerate}
\item
\begin{equation}
\label{Cond1}
E(N, \vec{Z}; \alpha ) < E(N, \vec{Z}, R = \infty; \alpha)
:= \lim_{R \rightarrow \infty} E(N, \vec{Z}, R; \alpha).
\end{equation}
\item 
The infimum is attained for some $R_0$, i.e.,
\begin{eqnarray}
\label{Cond2}
E(N, \vec{Z}; \alpha ) = E( N, \vec{Z}, R_0; \alpha ),
\end{eqnarray}
with the further requirement that it is an eigenvalue below the
essential spectrum for $H(N,\vec{Z},\vec{R};\alpha)$ if
$R = R_0$. 
\end{enumerate}
The first requirement ensures that the individual atoms stay bounded,
the second requirement ensures that all electrons remain bounded to
the molecule when the bond distance is $R_0$. We identify in the
physical context $R_0$ as an \textit{equilibrium bond distance}, i.e.,
the distance between $\vec{R}_1$, $\vec{R}_2$ in
a stable molecule. Note that we do not exclude the existence of more
than one $R_0$; this case plays no role in the present consideration
though.\\ 

If stability holds, by which we mean the existence of a stable
Born--Oppenheimer ground state as defined above, we introduce the
binding energy for a molecule as
\begin{equation}
\Delta E_{b}( N, \vec{Z}; \alpha ) = 
E( N, \vec{Z}, R = \infty; \alpha ) - 
E( N, \vec{Z}, R_0; \alpha ) > 0 .
\end{equation}

We state first the main results of this paper before establishing the
proofs in the subsequent sections. The final proofs of Theorem~1 and
Theorem~2 can be found in Sec.~V and Sec.~VI respectively while
Secs.~II-IV introduce necessary prerequisites.\\

\textbf{Theorem 1} \begin{slshape} 
Assume $H(N,\vec{Z},\vec{R};\alpha)$ has a stable ground
state for some $Z = Z_1 + Z_2$, with 
$\underset{k}{\mbox{max}} \, \{Z_k \alpha\} < 2/\pi$. 
Let $\varepsilon > 0$ be such that  
$\underset{k}{\mbox{max}} \, \{Z_k \alpha\} \le 
\frac{2}{\pi}(1-\varepsilon)$. Then,
independent of particle symmetry,
\begin{eqnarray}
\frac{Z_1 Z_2}{Z_1 + Z_2}
& \le & N \left( \frac{1}{2} + \sqrt{
\frac{1}{4} + 3 \, \sigma(\varepsilon,\tau)} \right) ,
\end{eqnarray}
with $\sigma(\varepsilon,\tau) = 2 \left[ 1 + \varepsilon^{-1} \,
(1-\varepsilon)^{-1} \right] \tau$, where $\tau > 0$ is the constant
in Lemma~3, (\ref{Lemma3}).
\end{slshape}\\

Assuming fermions, we obtain instead
\begin{eqnarray}
\frac{Z_1 Z_2}{Z_1 + Z_2}
& \le & N \, \left( \frac{1}{2} + \sqrt{
\frac{1}{4} + \frac{3 \sigma(\varepsilon,\tau)}{N^{2/3}} } \right) .
\end{eqnarray}\\

A consideration of the case $Z_1 \gg Z_2$ shows that the bound
on $N$ is actually controlled by the smaller of the
two atoms.\\

An upper bound on $N$ is given by Lieb\cite{Lieb84} as $N < 2 (Z_1 +
Z_2) + 2$, valid for the nonrelativistic as well as the relativistic
case. In the latter, Dall'Acqua, S{\o}rensen and
Stockmeyer\cite{DSS} proved this bound to hold for
$\underset{k}{\mbox{max}} \, \{Z_k \alpha\} < 2/\pi$.\\

\textbf{Theorem 2} \begin{slshape} There exist constants $c_1, c_2 >
  0$ such that if $H(N,\vec{Z},\vec{R};\alpha)$ has a
  stable ground state on the fermionic space for $Z = Z_1 + Z_2$, $Z =
  N$, with $\underset{k}{\mbox{max}} \, \{Z_k \alpha \} < 2/\pi$, $k =
  1,2$, then
\begin{eqnarray}
0 < \Delta E_b (\vec{Z};\alpha) < c_1 \, Z^{2-1/30} \\
R_0 \ge c_2 \, Z^{-1/3+11/210}.
\end{eqnarray}
\end{slshape}\\

From Thomas--Fermi theory we know that the bulk of electrons around
one nucleus is at a distance $Z^{-1/3}$ from that nucleus. Thus the
theorem states that the internuclear distance is much larger than the
radius of the bulk electron cloud.\\

The approach in the proof of Theorem~1 consists of the introduction of
localization functions $\chi_1, \chi_2 \in H^1
\left(\mathbb{R}^3 \times \mathbb{R}^3 \right)$ satisfying
\begin{equation}
\chi_1 (x, \vec{R})^2 + \chi_2 (x, \vec{R})^2
= 1. 
\end{equation}
The nonrelativistic terms $H_{ee}+H_{en}+H_{nn}$ can be treated
according to a localization argument following \cite{Sol90}. The
nonlocal operator $H^{\alpha}$ will be treated separately, applying
here results from \cite{Sol10}. We will begin with this consideration
in Sec.~\ref{sec:IMS}, before introducing the localization argument in
Sec.~\ref{sec:local}. In Sec. \ref{sec:R0} we establish an estimate on
$R_0$ as a function of $N$. We introduce in this context Lemma~3 and
Theorem~4, the proofs of which can be found in the appendix. These
results will be used in Sec.~\ref{sec:T1} to prove Theorem~1, i.e., to
derive a global bound on $N$. Theorem~2 will be proved in
Sec.~\ref{sec:T2}, where we will apply the recently proved
relativistic Scott correction\cite{Sol10} to
derive bounds on $\Delta E_b$, $R_0$.\\

We will in the following denote the wave function of the many-particle
system by $\Psi$, with $\Psi \in \bigwedge^N L^2(\mathbb{R}^3 \times \{
-1,+1\})$, and $\| \Psi \|_2^2 = 1$.

\section{Relativistic IMS formula}
\label{sec:IMS}

We will first consider the kinetic energy operator $H^{\alpha} =
\sum\limits_{i} T_i^{\alpha}$ separately by application of the
relativistic Ismagilov--Morgan--Simon (IMS) formula. This formula was
proved in \cite{Sol10} generally for a family $( \xi_u)_{u \in
  \mathcal{M}}$ of positive bounded $C^1$-functions on $\mathbb{R}^3$
with bounded derivatives which, given a positive measure $d \mu$ on
$\mathcal{M}$, fulfill $\int_{\mathcal{M}} \xi_u(x)^2 d \mu(u) = 1$
for all $x \in \mathbb{R}^3$. We consider in the present context the
set of two localization functions $\chi_1$, $\chi_2$. The formula
reduces in this case, for any $\psi \in H^{1/2}(\mathbb{R}^3)$, to
\begin{eqnarray}
\langle \psi | \left( \sqrt{- \alpha^{-2 } \, \Delta + \alpha^{-4}}
- \alpha^{-2} \right) | \psi \rangle & = & \langle \psi | T^{\alpha}
| \psi \rangle \nonumber \\
& = &\langle \psi | \chi_1 \, T^{\alpha} \, \chi_1 | \psi \rangle - 
\langle \psi | L_{\chi_1} | \psi \rangle \nonumber \\
 && + \langle \psi | \chi_2 \, T^{\alpha} \, \chi_2 | \psi \rangle - 
\langle \psi | L_{\chi_2} | \psi \rangle,
\end{eqnarray}
with $L_{\chi_k}$ denoting an operator with integral kernel
\begin{eqnarray}
L_{\chi_k}(x,y) = (2 \pi)^{-2} \, \alpha^{-3} \, |x - y|^{-2} \:
K_2\left( \alpha^{-1} \, |x-y| \right) \: \left[ \chi_k(x) - \chi_k(y) \right]^2,
\end{eqnarray}
where $K_2$ is a modified Bessel function defined by
\begin{eqnarray}
K_2(t) = t \int_0^{\infty} e^{-t \sqrt{s^2 + 1}} s^2 ds.
\end{eqnarray}
Note that our choice of unit system differs from \cite{Sol10},
Theorem~13 (\textsl{Relativistic IMS formula)}, in setting $m =
1$.\\

Employing the relativistic IMS formula, we can reformulate the
following matrix element as
\begin{eqnarray}
\label{Hrel1}
\langle \Psi | H^{\alpha} | \Psi \rangle
& = & \sum_i \Big\{ \langle \Psi | \chi_1 \, T_i^{\alpha} 
\, \chi_1 | \Psi \rangle
+ \langle \Psi | \chi_2 \, T_i^{\alpha} \, 
\chi_2 | \Psi \rangle \nonumber \\
&& - \langle \Psi | L_{\chi_1}(x_i,y_i) | \Psi \rangle
- \langle \Psi | L_{\chi_2}(x_i,y_i) | \Psi \rangle \Big\} .
\end{eqnarray}

\section{Localization Argument}
\label{sec:local}

We consider now two-cluster decompositions $\theta = ( \theta_1,
\theta_2)$ of $\{ 1, \: \ldots, \: N\}$. The interactions within the
two separate clusters are described by $H_{\theta_1}$, $H_{\theta_2}$,
with $H_{\theta} = H_{\theta_1} + H_{\theta_2}$ and
\begin{eqnarray}
\label{Htheta}
H_{\theta_k} = \sum_{i \in \theta_k} \left( \sqrt{-
    \alpha^{-2} \, \Delta_i
+ \alpha^{-4}} - \alpha^{-2} \right) 
- \sum_{i \in \theta_k}  \frac{Z_k}{\left| x_i - \vec{R}_k \right|} 
+ \sum_{\genfrac{}{}{0.pt}{2}{i<j}{i,j \in \theta_k}} \frac{1}{\left| x_i - x_j \right|},
\end{eqnarray}
while the intercluster potential $I_{\theta}$ can be given as
\begin{eqnarray}
\label{Itheta}
I_{\theta} = - \sum_{i \in \theta_2} 
\frac{Z_1}{\left| x_i - \vec{R}_1 \right|}
- \sum_{i \in \theta_1} \frac{Z_2}{\left| x_i - 
\vec{R}_2 \right|}
+ \sum_{i \in \theta_1 \atop j \in \theta_2} 
\frac{1}{\left| x_i - x_j \right|}
+ \frac{Z_1 Z_2}{R}, 
\end{eqnarray}
and we recognize that $H = H_{\theta} + I_{\theta}$. \\

The space of localization functions can be correspondingly regrouped as
\begin{eqnarray*}
\prod_{i=1}^N \left( \chi_1 (x_i)^2 + \chi_2 (x_i)^2 \right)
& = & \sum_{\theta} \prod_{i \in \theta_1} 
\chi_1 (x_i)^2 \prod_{j \in \theta_2} 
\chi_2 (x_j)^2 = \sum_{\theta} \chi_{\theta}^2
\end{eqnarray*}
with $\chi_{\theta} = \prod\limits_{i \in \theta_1} \chi_1
(x_i) \prod\limits_{j \in \theta_2} \chi_2 (x_j)$,
omitting here and from now on out of reasons of simplicity the
dependence in $\vec{R}$.\\

We consider, starting from (\ref{Hrel1}), that
\begin{eqnarray*}
\langle \Psi | H^{\alpha} | \Psi \rangle
& = & \sum_i \Big\{ \langle \Psi | \chi_1 (x_i) \, T_i^{\alpha} 
\, \chi_1 (x_i) \prod_{j \neq i}^N  
\left( \chi_1 (x_j)^2 + \chi_2 (x_j)^2 \right) | \Psi \rangle \\
& & + \langle \Psi | \chi_2 (x_i) \, T_i^{\alpha} 
\, \chi_2 (x_i) \prod_{j \neq i}^N  
\left( \chi_1 (x_j)^2 + \chi_2 (x_j)^2 \right)
| \Psi \rangle \Big\} \\
& & - \sum_i \Big\{\langle \Psi | L_{\chi_1} (x_i, y_i) | \Psi \rangle
+ \langle \Psi | L_{\chi_2} (x_i, y_i) | \Psi \rangle \Big\} \\
& = & \sum_{\theta} \Big\{ \langle \Psi | \chi_{\theta}
\sum_i T_i^{\alpha} \chi_{\theta} | \Psi \rangle \Big\} \\
& & - \sum_i \Big\{\langle \Psi | L_{\chi_1} (x_i, y_i) | \Psi \rangle
+ \langle \Psi | L_{\chi_2} (x_i, y_i) | \Psi \rangle \Big\},
\end{eqnarray*}
and furthermore reformulate the Coulomb terms straightforwardly
\begin{eqnarray*}
\langle \Psi | (H-H^{\alpha}) | \Psi \rangle
& = & \sum_{\theta} \langle \Psi | (H-H^{\alpha})
\chi_{\theta}^2 | \Psi \rangle \\
& = & \sum_{\theta} \langle \Psi | \chi_{\theta} \, (H-H^{\alpha})
\chi_{\theta} | \Psi \rangle .
\end{eqnarray*}

Additionally, we recall that if $H$ has a stable ground state then
\begin{eqnarray*}
E(N, \vec{Z}; \alpha) & = & \inf \: \mbox{spec} \: 
 H(N,\vec{Z},\vec{R};\alpha)\big|_{R = R_0} < 
E_{\theta}(N,\vec{Z},R_0;\alpha) \\
& = & \inf \: \mbox{spec} \: 
 H_{\theta}(N,\vec{Z},\vec{R};\alpha)\big|_{R = R_0}.
\end{eqnarray*}
This is the Hunziker--van Winter--Zhislin (HVZ) theorem\cite{CycSG},
in its formulation for this operator.\\

We can therefore estimate
\begin{eqnarray}
R \langle \Psi | H | \Psi \rangle
& = & R \sum_{\theta} \Big\{ \langle \Psi | \chi_{\theta}
\, H^{\alpha} \, \chi_{\theta} | \Psi \rangle
+ \langle \Psi | \chi_{\theta} \, (H-H^{\alpha}) \,
\chi_{\theta} | \Psi \rangle \Big\} \nonumber \\
&& -  R \sum_i \Big\{\langle \Psi | L_{\chi_1} (x_i, y_i) | \Psi \rangle
+ \langle \Psi | L_{\chi_2} (x_i, y_i) | \Psi \rangle \Big\} \nonumber \\
& = & R \sum_{\theta} \Big\{ \langle \Psi | \chi_{\theta} \,
H_{\theta} \, \chi_{\theta} | \Psi \rangle
+ R \langle \Psi | I_{\theta} \,
\chi_{\theta}^2 | \Psi \rangle \Big\} \nonumber \\
&& -  R \sum_i \Big\{\langle \Psi | L_{\chi_1} (x_i, y_i) | \Psi \rangle
+ \langle \Psi | L_{\chi_2} (x_i, y_i) | \Psi \rangle \Big\} \nonumber \\
& \ge & R_0 \sum_{\theta} \Big\{ E_{\theta} 
\langle \Psi | \chi_{\theta}^2 | \Psi \rangle
+ \langle \Psi | I_{\theta} \, \chi_{\theta}^2 | \Psi \rangle
\Big\} \nonumber \\
&&  -  R_0 \sum_i \Big\{\langle \Psi | L_{\chi_1} (x_i, y_i) | \Psi \rangle
+ \langle \Psi | L_{\chi_2} (x_i, y_i)  | \Psi \rangle \Big\} \nonumber \\
& \ge & R_0 \, \inf_{\theta} E_{\theta}
+ R_0 \sum_{\theta} \langle \Psi | I_{\theta} \chi_{\theta}^2 |
\Psi \rangle \nonumber \\
&& - R_0 \sum_i \Big\{\langle \Psi | L_{\chi_1} (x_i, y_i) | \Psi \rangle
+ \langle \Psi | L_{\chi_2} (x_i, y_i) | \Psi \rangle \Big\}.
\end{eqnarray}

We obtain a bound on the localization error $\langle \psi |
L_{\chi_k} (x_i, y_i)  | \psi \rangle$ in the following as
\begin{eqnarray}
\left| \langle \psi | L_{\chi_k} (x, y) | \psi \rangle \right|
& \le & (2 \pi)^{-2} \alpha^{-3} \int | \psi(x) | \,
| \psi(y) | \; K_2\left( \alpha^{-1} \, |x-y| \right) 
\frac{| \chi_k(x) - \chi_k(y)|^2}{|x-y|^2} dx \, dy \nonumber \\
& \le & (2 \pi)^{-2} \alpha^{-3} \, \parallel \nabla \chi_k \parallel^2_{\infty}
 \int | \psi(x) | \, | \psi(y) | \; 
K_2\left( \alpha^{-1} \, |x-y| \right) dx \, dy \nonumber \\
& \le & (2 \pi)^{-2} \alpha^{-3} \, \parallel \nabla \chi_k \parallel^2_{\infty}
\int | \psi(x) |^2 \: K_2\left( \alpha^{-1} \, |x-y| \right) dx \, dy \nonumber \\
& = & (2 \pi)^{-2} \, \parallel \nabla \chi_k \parallel^2_{\infty}
\int | \psi(x) |^2 dx \int K_2 \left( | y | \right) dy \nonumber \\
\label{LErr1}
& = & \frac{3}{2} \, \parallel \nabla \chi_k \parallel^2_{\infty}
\| \psi \|_2^2,
\end{eqnarray}
where in the third step we used the Cauchy-Schwarz inequality and
$K_2 \ge 0$. We obtain from this the final estimate
\begin{eqnarray}
- \Delta E_b ( N, \vec{Z}; \alpha ) \, R_0 & \ge &
- \frac{3}{2} \, N \, R_0 \, \left\{ 
\parallel \nabla \chi_1 \parallel^2_{\infty}
+ \parallel \nabla \chi_2 \parallel^2_{\infty} \right\} \nonumber \\
&& - R_0 \sum_{i=1}^N \langle \Psi | \frac{Z_1}{|x_i - \vec{R}_1|}
\chi_2(x_i)^2 + \frac{Z_2}{|x_i - \vec{R}_2|}
\chi_1(x_i)^2 | \Psi \rangle \nonumber \\
&& + R_0 \sum_{i < j}^N \langle \Psi | \frac{1}{|x_i-x_j|}
\left( \chi_1(x_i)^2 \, \chi_2(x_j)^2 +
\chi_1(x_j)^2 \, \chi_2(x_i)^2 \right) | \Psi \rangle 
\nonumber \\
\label{Eb1}
&& + Z_1 \, Z_2 .
\end{eqnarray}

\section{Estimate on $R_0$}
\label{sec:R0}

Following \cite{Sol90} we introduce explicit expressions for the
localization functions which exhibit a dependence on the parameter
$\mu = Z_2/Z_1$. We impose for convenience $Z_1 \ge Z_2$,
therefore $\mu \le 1$. Introducing
\begin{eqnarray*}
\overline{x}=x+\frac{1-\mu}{2(\mu + 1)}(\vec{R}_1-\vec{R}_2) \quad \mbox{and} \quad
\overline{\vec{R}}=\frac{1}{\mu + 1}(\vec{R}_1-\vec{R}_2)
\end{eqnarray*}
we obtain
\begin{eqnarray*}
\overline{x}+\mu \vec{\overline{R}}=x-\vec{R}_2
\quad \mbox{and} \quad
\overline{x}-\vec{\overline{R}}=x-\vec{R}_1,
\end{eqnarray*}
using that $\vec{R}_2 = -\vec{R}_1$.\\

We define the localizing functions as
\begin{eqnarray*}
\chi_1(x) = \frac{|\overline{x}+\mu \vec{\overline{R}}|}{\sqrt{\mu+1}
(|\overline{x}|^2+\mu \overline{R}^2)^{1/2}} \\ 
\chi_2(x) = \frac{\sqrt{\mu} |\overline{x}-\vec{\overline{R}}|}
{\sqrt{\mu+1}(|\overline{x}|^2+\mu \overline{R}^2)^{1/2}},
\end{eqnarray*}
with $\overline{R} = |\overline{\vec{R}}|$. It is straightforward to
see that 
\begin{eqnarray*}
\chi_1(x)^2 + \chi_2(x)^2 = 1.
\end{eqnarray*}

We evaluate 
\begin{eqnarray}
R \, \left( (\nabla \chi_1(x))^2 + (\nabla \chi_2(x))^2 \right)
& = & \frac{\mu}{(\mu + 1)^2} \frac{R^3}{(|\overline{x}|^2+\mu
  \overline{R}^2)^2} \nonumber \\
& \le & \frac{(\mu + 1)^2}{\mu} R^{-1}.
\end{eqnarray}

We will use this expression for $R$ being the equilibrium bond
distance $R_0$. We recognize that an estimate on $R_0^{-1}$ is
necessary to subsequently deduce an estimate on the localization
error. Obtaining this estimate is more involved in the relativistic
case than in the nonrelativistic one due to the presence of the
critical point if $Z_k \alpha = \frac{2}{\pi}$ for some $k$. We start
out by comparing the energy $E$ of the diatomic system with the
situation of localizing all electrons on one of the centers, say
$\vec{R}_1$. We obtain using the stability conditions (\ref{Cond1}),
(\ref{Cond2}) that for the energy $E^{at}$ of a single atom
\begin{eqnarray}
E^{at}(N,Z_1;\alpha) & \ge &
\lim_{R \rightarrow \infty} E(N,\vec{Z},R;\alpha) \nonumber \\
& \ge & E(N,\vec{Z},R_0;\alpha) \nonumber \\
\label{boundUA}
& \ge & E^{at}(N,Z_1+Z_2;\alpha) + \frac{Z_1 Z_2}{R_0}.
\end{eqnarray}
The last inequality, the lower bound on the energy of a united atom,
is simple to derive for the nonrelativistic case, as was done in
\cite{ThirMPIII} (Sec. $(4.6.14)$). The extension to the present case
of the relativistic operator is straightforward as the form of the
one-particle operator $H^{\alpha}$ is identical for the molecular case
and the united atom. \\

Establishing an upper bound on $E^{at}(N,Z_1;\alpha) -
E^{at}(N,Z_1+Z_2;\alpha)$ will allow us to
deduce an upper bound on $R_0^{-1}$. \\

Since $Z \mapsto E^{at}(N,Z;\alpha)$ is a nonincreasing, concave function we
have
\begin{eqnarray*}
-\frac{E^{at}(N,Z_1;\alpha)-E^{at}(N,Z_1+Z_2;\alpha)}{Z_2} \ge 
\left[ \frac{\partial E^{at}(N,Z_1+Z_2;\alpha)}{\partial Z} \right]_-,
\end{eqnarray*}
where $\left[ \frac{\partial}{\partial Z} \cdot \right]_-$ refers to the
left derivate.\\

We note that if
\begin{eqnarray*}
\left[ \frac{\partial E^{at}(N,Z;\alpha)}{\partial Z} \right]_- = 0
\end{eqnarray*}
we trivially have a lower bound of zero while if
\begin{eqnarray*}
\left[ \frac{\partial E^{at}(N,Z;\alpha)}{\partial Z} \right]_- < 0,
\end{eqnarray*}
then $E^{at}(N,Z;\alpha) < 0$. In this case there is an $1 \le n \le N$ such
that $E^{at}(N,Z;\alpha) = E^{at}(n,Z;\alpha) < E^{at}(n-1,Z;\alpha)$
and there is an $n$-particle wave function $\Psi_n$ such that
\begin{eqnarray*}
H^{at}(n,Z;\alpha) \, \Psi_n = E^{at}(n,Z;\alpha) \, \Psi_n ,
\end{eqnarray*}
with $H^{at}(n,Z;\alpha)$ being the Hamiltonian of the atomic
system.\\

By the Feynman-Hellman Theorem we have 
\begin{eqnarray*}
\left[ \frac{\partial E^{at}(N,Z;\alpha)}{\partial Z} \right]_-
\ge - \left( \Psi_n, \sum_{i=1}^n \frac{1}{|x_i|} \Psi_n \right),
\end{eqnarray*}
and have therefore reformulated the problem into establishing an upper
bound on $\left( \Psi_n, \sum_{i=1}^n \frac{1}{|x_i|} \Psi_n \right)$
when $E^{at}(n,Z;\alpha) < 0$.\\

This we will achieve in the following by first providing a lower bound
on 
\begin{eqnarray*}
\sum_{i=1}^n \left[
\sqrt{-\alpha^{-2} \, \Delta_i + \alpha^{-4}}
- \alpha^{-2} - \frac{Z}{|x_i|} \right] = 
\sum_{i=1}^n \left[ T_i^{\alpha} - \frac{Z}{|x_i|} \right]
\end{eqnarray*}
for $Z \alpha \le \frac{2}{\pi}$ (Lemma~3, (\ref{Lemma3})) which will
be accomplished by use of the combined Daubechies--Lieb--Yau (DLY)
inequality (see \cite{Sol10}, furthermore \cite{Daub83b,Lieb88} for
Daubechies inequality and Lieb--Yau inequality). This lower bound will
then be used in the proof of Theorem~4, namely the upper bound on
$\left( \Psi_n, \sum_{i=1}^n \frac{1}{|x_i|} \Psi_n \right)$, for the
case of $Z \alpha \le \frac{2}{\pi} \, (1 - \varepsilon)$. After
having established Theorem~4, we are in the position to return to
(\ref{boundUA}) and provide a bound on $R_0^{-1}$. We recognize that
our approach fails for
the critical case of $Z \alpha = \frac{2}{\pi}$.\\

\textbf{Lemma 3} \begin{slshape} If $Z \alpha \le \frac{2}{\pi}$
  then for all $n \ge 1$
\begin{eqnarray}
\label{Lemma3}
\sum_{i=1}^n \left[ T_i^{\alpha} - \frac{Z}{|x_i|} \right]
\ge - Z^2 \, \tau \left\{ \begin{array}{l@{\quad}l}
n & \mbox{without particle symmetry} \\
n^{1/3} & \mbox{assuming fermions}
\end{array} \right.
= - Z^2 \, \kappa(n),
\end{eqnarray}
for some constant $\tau > 0$, with the function $\kappa(n)$ defined by
(\ref{Lemma3}).
\end{slshape}\\

\textbf{Theorem 4} \begin{slshape} Assume $Z \alpha \le
  \frac{2}{\pi}(1-\varepsilon)$, $0 < \varepsilon < 1$. Assume
  $\Psi_n$ eigenfunction of $H^{at}(n,Z;\alpha)$ with eigenvalue
  $E^{at}(n,Z;\alpha)$, where $1 \le n \le N$ and $E^{at}(N,Z;\alpha) =
  E^{at}(n,Z;\alpha) < E^{at}(n-1,Z;\alpha)$. Then
\begin{eqnarray}
\left[ \frac{\partial E^{at}(N,Z;\alpha)}{\partial Z} \right]_- \ge
- \left( \Psi_n, \sum_{i=1}^n \frac{1}{|x_i|} \Psi_n \right)
\ge - \left[ 1 + \varepsilon^{-1} \, (1-\varepsilon)^{-1}
\right] \, \kappa(n) \, Z.
\end{eqnarray}
\end{slshape}\\

The proofs of Lemma~3 and Theorem~4 can be found in the appendix.\\

We are now in the position to obtain our final estimate on
$R_0^{-1}$. Starting with the bound on the united atom we reformulate
\begin{eqnarray*}
\frac{Z_1 Z_2}{R_0} && \le E^{at}(N,Z_1;\alpha) - E^{at}(N,Z_1+Z_2;\alpha) \\
&& \le -Z_2 \left[ \frac{\partial E^{at}(N,Z_1+Z_2;\alpha)}
{\partial Z} \right]_- \\
&& \le Z_2 \, (Z_1 + Z_2) \left[ 1 + \varepsilon^{-1} \, (1-\varepsilon)^{-1}
\right] \, \kappa(N) \\
&& \le 2 \left[ 1 + \varepsilon^{-1} \, (1-\varepsilon)^{-1}
\right] \, \kappa(N) \, Z_1Z_2 ,
\end{eqnarray*}
with $Z_2 \le Z_1$. We have used that $n \le N$.\\

We therefore obtain
\begin{eqnarray}
\label{Rbound}
R_0^{-1} \le \sigma(\varepsilon,\tau) \,
\left\{ \begin{array}{l@{\quad}l}
N & \mbox{without particle symmetry} \\
N^{1/3} & \mbox{assuming fermions} ,
\end{array} \right.
\end{eqnarray}
where $\sigma(\varepsilon,\tau) = 2 \left[ 1 + \varepsilon^{-1} \,
  (1-\varepsilon)^{-1} \right] \tau$, for some constants $1 >
\varepsilon > 0$, $\tau > 0$, with $\tau$ introduced in Lemma~3,
(\ref{Lemma3}).

\section{Global Bound on $N$}
\label{sec:T1}

The estimate on $R_0$ as a function of $N$ allows us to estimate the
localization error from (\ref{LErr1}) as 
\begin{eqnarray}
R_0 \left| \langle \psi | L_{\chi_1} (x, y) | \psi \rangle \right|
+ R_0 \left| \langle \psi | L_{\chi_2} (x, y) | \psi \rangle \right|
& \le & \frac{3}{2} \, R_0 \, \left\{ \parallel \nabla \chi_1 \parallel^2_{\infty}
+ \parallel \nabla \chi_2 \parallel^2_{\infty} \right\} \nonumber \\
& \le & 3 R_0 \, \parallel (\nabla \chi_1)^2 + (\nabla \chi_2)^2
  \parallel_{\infty} \nonumber \\
\label{LErr2}
& \le & 3 \, \sigma(\varepsilon,\tau) \, N \, 
\frac{(\mu + 1)^2}{\mu}.
\end{eqnarray}

Inspecting the remaining terms in (\ref{Eb1}), we recognize the
difficulty in obtaining an estimate on the repulsion of electrons
between different clusters and will neglect the term at this point. We
thereby obtain from (\ref{Eb1})
\begin{eqnarray}
\label{Est1}
0 \ge - 3 \, \sigma(\varepsilon,\tau)
\frac{(\mu + 1)^2}{\mu} N^2
- R_0 \sum_{i=1}^N \langle \Psi |
\frac{Z_2 |\overline{x}_i+\mu \vec{\overline{R}}|+
\mu Z_1 |\overline{x}_i- \vec{\overline{R}}|
}{(\mu + 1)(|\overline{x}_i|^2+\overline{R}^2)}|\Psi \rangle
+ Z_1 Z_2 .
\end{eqnarray}

We estimate following \cite{Sol90}
\begin{eqnarray}
\label{Est_en}
\left( Z_2 |\overline{x}+\mu \vec{\overline{R}}|
+ \mu Z_1 |\overline{x}-\vec{\overline{R}}| \right)^2
& \le & (Z_2^2+Z_1^2 \mu) \left( |\overline{x}+\mu \vec{\overline{R}}|^2
+ |\overline{x}-\vec{\overline{R}}|^2 \right) \\
& = & (Z_2^2+Z_1^2 \mu) (\mu + 1)
( |\overline{x}|^2 + \mu \overline{R}^2 ) \nonumber
\end{eqnarray}
and obtain with $\mu = Z_2/Z_1$
\begin{eqnarray*}
0 \ge - 3 \, \sigma(\varepsilon,\tau) \,
N^2 - N \frac{Z_1 Z_2}{Z_1 + Z_2} + 
\left( \frac{Z_1 Z_2}{Z_1 + Z_2} \right)^2,
\end{eqnarray*}
therefore
\begin{eqnarray*}
\frac{Z_1 Z_2}{Z_1 + Z_2}
& \le & N \left( \frac{1}{2} + \sqrt{
\frac{1}{4} + 3 \, \sigma(\varepsilon,\tau)} \right) .
\end{eqnarray*}

We compare this result to the nonrelativistic case where
Solovej\cite{Sol90} proved
\begin{eqnarray*}
\frac{Z_1 Z_2}{Z_1 + Z_2}
& \le & N \, (3/2).
\end{eqnarray*}
The difference in the bound on $N$ is attributable to the
prefactor in the first term of (\ref{Est1}) which is in
the nonrelativistic case $-\frac{3}{4}$ instead of $-3
\sigma(\varepsilon,\tau)$. Aside of the more demanding assessment of
$R_0^{-1}$, which in the nonrelativistic case yields an estimate of
$R_0^{-1} \le \frac{3}{2} \, N$, we recognize two further steps in the
present derivation which influence our bound. The simple estimate of
$\parallel \nabla \chi_1 \parallel^2_{\infty} + 
\parallel \nabla \chi_2 \parallel^2_{\infty} \, \le 2
\, \parallel (\nabla \chi_1)^2 + (\nabla \chi_2)^2 \parallel_{\infty}$ in the second
step of (\ref{LErr2}) introduces a factor of two. A further factor
of three has to be attributed to the estimates in
(\ref{LErr1}). In particular, we bound the expression $| \chi_k(x)
- \chi_k(y)|$ by
\begin{eqnarray*}
| \chi_k(x) - \chi_k(y)| \le \parallel \nabla \chi_k \parallel_{\infty} \cdot |x-y|
\end{eqnarray*}
before employing subsequently the Cauchy-Schwarz inequality. Instead,
it would be preferable to construct explicit forms for $\chi_1$,
$\chi_2$ which would allow a tighter estimate of $\langle \psi |
L_{\chi_k}| \psi \rangle$ while at the same time
permitting an advantageous evaluation of the electron--nuclear
attraction term in (\ref{Est_en}). With respect to the
localization error, one has in particular to construct a form of
localization functions which results in a cancellation of the
denominator $|x-y|$ in order to obtain a bound of $| \langle \psi |
L_{\chi_k}| \psi \rangle | < \infty$. In the present
context, we have decided to employ the particular forms of $\chi_1$,
$\chi_2$ given in \cite{Sol90} which result in an algebraically simple
expression of the bound on $N$ and recognize the scope that exists in
the possible improvement on the bound of the localization error.\\

Considering alternatively the fermionic bound on $R_0^{-1}$
in our estimate, we obtain 
\begin{eqnarray*}
\frac{Z_1 Z_2}{Z_1 + Z_2}
& \le & N \, \left( \frac{1}{2} + \sqrt{
\frac{1}{4} + \frac{3 \sigma(\varepsilon,\tau)}{N^{2/3}} } \right) .
\end{eqnarray*}
This can be compared to Benguria, Siedentop and Stockmeyer\cite{Ben01}
who find a very similar expression for the case of homonuclear
relativistic molecular ions. Their proof holds for $Z \alpha \le 1/2$,
where $Z = Z_1 + Z_2 = 2 Z_1$.

\section{Bound on $\Delta E_b$, $R_0$}
\label{sec:T2}

The first rigorous investigaton of the limit $Z_k \rightarrow \infty$
with $Z_k \alpha$ bounded was given by S{\o}rensen\cite{Thom05}. The
leading asymptotics of the ground state were established to be
determined by Thomas--Fermi theory. The first correction term, the
Scott correction, and its dependence on $Z_k \alpha$, was
proved by Solovej, S{\o}rensen and Spitzer\cite{Sol10} for a neutral
molecular system, i.e., $N = Z = \sum_1^M Z_k$, in the framework of
the Born-Oppenheimer formulation. \\

We state their main theorem, the relativistic Scott correction, for
the case of a diatomic system. Let $Z = Z_1 + Z_2$, $\vec{z} =
(z_1,z_2) = Z^{-1} \vec{Z}$ with $z_1, z_2 > 0$, thereby $z_1 + z_2 =
1$. We define $\vec{r} = Z^{1/3} \vec{R}$, with $|\vec{r}| > r_0$ for
some $r_0 > 0$. Note that $\vec{r} = \vec{r}_1 - \vec{r}_2$, with
$\vec{r}_1, \vec{r}_2 \in \mathbb{R}^3$, furthermore $r = |\vec{r}_1 -
\vec{r}_2|$. Then there exist a value $E^{TF}(\vec{z},r)$ and a
universal (independent of $\vec{z}$ and $r$) continuous,
nonincreasing function $\mathcal{S} \, : \, [0,2/\pi] \rightarrow
\mathbb{R}$ with $\mathcal{S}(0) = 1/4$ such that as $Z \rightarrow
\infty$ and $\alpha \rightarrow 0$, with $\underset{k}{\mbox{max}} \,
\{Z_k \alpha\} \le 2/\pi$, we have
\begin{eqnarray}
\label{Scott}
E(\vec{Z},R;\alpha) = Z^{7/3} \, E^{TF}(\vec{z},r)
+ 2 \, Z_1^2 \, \mathcal{S}(Z_1 \alpha)
+ 2 \, Z_2^2 \, \mathcal{S}(Z_2 \alpha)
+ \mathcal{O}(Z^{2-1/30}).
\end{eqnarray}
The error term means that $|\mathcal{O}(Z^{2-1/30})| < c \,
Z^{2-1/30}$, where the constant $c$ only depends on $r_0$. 
Moreover, the Thomas--Fermi energy satisfies the scaling relation
$E^{TF}(\vec{Z},R) = Z^{7/3} E^{TF}(\vec{z},r)$. We call
$E^{TF}(\vec{Z},R)$ the Thomas--Fermi energy of the molecule.\\

We consider from (\ref{Rbound}) that
\begin{eqnarray*}
R_0 \ge \frac{1}{\sigma(\varepsilon,\tau)} N^{-1/3} 
= \frac{1}{\sigma(\varepsilon,\tau)} \, Z^{-1/3},
\end{eqnarray*}
therefore asserting that $R_0 > Z^{-1/3} r_0$ for some constant
$r_0 > 0$.\\

We will use in the following evaluation scaling properties from
Thomas--Fermi theory. In particular, it is established (see
\cite{Lieb81} for a review of TF-theory) that
\begin{eqnarray}
\label{TF_SP}
E^{TF}(\vec{Z},R) = Z^{7/3} E^{TF}(\vec{z},r).
\end{eqnarray}

We introduce furthermore the estimate on the split of a diatomic
system in Thomas--Fermi theory which was derived in \cite{Sol90}
based on Brezis--Lieb\cite{Brez79}
\begin{eqnarray}
\label{TF_NS}
E^{TF}(\vec{z},r) \ge E^{TF}(z_1) +  E^{TF}(z_2) + c \, r^{-7},
\end{eqnarray}
with some constant $c$. $E^{TF}(z_n)$ denotes the scaled
Thomas--Fermi energy of a nucleus of charge $n$ at the origin.\\

We consider now the case of a diatomic molecule and obtain by
application of the relativistic Scott correction (\ref{Scott})
and furthermore (\ref{TF_SP}), (\ref{TF_NS}) that
\begin{eqnarray}
H(N,\vec{Z},\vec{R};\alpha)
& \ge & E(\vec{Z},R;\alpha) \nonumber \\
& \ge & Z^{7/3} E^{TF}(z_1) + 2 \, Z_1^2 \, \mathcal{S}(Z_1 \alpha)
+ Z^{7/3} E^{TF}(z_2) + 2 \, Z_2^2 \, \mathcal{S}(Z_2 \alpha)
\nonumber \\
& & + c_1 \, r^{-7} \, Z^{7/3} - c_0 \, Z^{2-1/30} \nonumber \\
& = & E^{TF}(Z z_1) + 2 \, Z_1^2 \, \mathcal{S}(Z_1 \alpha)
+ E^{TF}(Z z_2) + 2 \, Z_2^2 \, \mathcal{S}(Z_2 \alpha)
\nonumber \\
& & + c_1 \, r^{-7} \, Z^{7/3} - c_0 \, Z^{2-1/30} \nonumber \\
& = & Z_1^{7/3} E^{TF}(1) + 2 \, Z_1^2 \, \mathcal{S}(Z_1 \alpha) 
+ Z_2^{7/3} E^{TF}(1) + 2 \, Z_2^2 \, \mathcal{S}(Z_2 \alpha) 
\nonumber \\
& & + c_1 \, r^{-7} \, Z^{7/3} - c_0 \, Z^{2-1/30} \nonumber \\
& \ge & E(Z_1;\alpha) + E(Z_2;\alpha) + 
 c_1 \, r^{-7} \, Z^{7/3} - c_0 \, Z^{2-1/30},
\end{eqnarray}
with constants $c_0$, $c_1$ (note that the definition of $c_0$ changes
in the last step).\\

We conclude
\begin{eqnarray}
E(N,\vec{Z},R;\alpha)
- E(Z_1;\alpha) - E(Z_2;\alpha) 
& \ge & c_1 \, R_0^{-7} - c_0 \, Z^{2-1/30},
\end{eqnarray}
and can therefore estimate
\begin{eqnarray}
0 & \ge & c_1 \, R_0^{-7} - c_0 \, Z^{2-1/30} \nonumber \\
R_0 & \ge & c \, Z^{-1/3+11/210}
\end{eqnarray}
The constant $c$ is independent of $Z$ and $R$ and may assume
different values in different inequalities.\\

We compare to the nonrelativistic case where Solovej\cite{Sol90}
proved
\begin{eqnarray*}
R_0 > c \, Z^{-(1/3)(1-\varepsilon)},
\end{eqnarray*}
for $N \le Z$ and $\varepsilon = 1/70$. \\

We note that the reason for our bound to be better than the bound of
Solovej is that we used that the energy is known up to the Scott
correction. The same can be done in the nonrelativistic case and
would give a similar bound. The nonrelativistic Scott correction was
proved by Ivrii and Sigal\cite{Ivri93} after the publication of
\cite{Sol90}.\\

Benguria, Siedentop and Stockmeyer\cite{Ben01} state for homonuclear
relativistic molecular ions a bound on $R_0$ (for $Z \alpha < 1/2$)
which, translated into our unit system, corresponds to an error of the
order $Z^{-1}$. The difference to our estimate reflects their
consideration of boltzonic electrons. \\

Recalling that in the case of a stable ground state
\begin{eqnarray*}
\Delta E_b (N,\vec{Z};\alpha) 
< E(Z_1;\alpha) + E(Z_2;\alpha) 
- E(N,\vec{Z},R;\alpha),
\end{eqnarray*}
we deduce furthermore a bound on $\Delta E_b (N,\vec{Z};\alpha)$ of
\begin{eqnarray}
\Delta E_b (N,\vec{Z};\alpha) < c \, Z^{2-1/30}.
\end{eqnarray}

\section{Acknowledgement}

I would like to thank Jan Philip Solovej for providing the framework
for this work as well as for numerous insightful and stimulating
discussions. \\

This work was supported by a fellowship within the Postdoc--Program of
the German Academic Exchange Service (DAAD).

\appendix
\section{Proofs}
\subsection{Proof of Lemma 3}

We begin with a separation of the nucleus--electron attraction by
considering a ball $|x| < r$ for some $r$ to be chosen later. The
contribution from the outside region is estimated by
$-\frac{Z}{r}$. We assume $n \ge 1$.
\begin{eqnarray}
\sum_{i=1}^n \left[ T_i^{\alpha} - \frac{Z}{|x_i|} \right]
& \ge & \sum_{i=1}^n \left[ T_i^{\alpha} - \frac{Z}{|x_i|}
  {\bf 1}_{\{|x_i|<r\}} \right]
- \frac{nZ}{r},
\end{eqnarray}
where ${\bf 1}_{\{|x|<r\}}$ is the characteristic function of the
ball $\{|x|<r\}$.\\

A lower bound on the right-hand side is in the fermionic case obtained
through a sum over all negative eigenvalues. In the case of no
particle symmetry, a lower bound is obtained by approximating the
bosonic states through $n$ times the lowest eigenvalue, which is then
in turn estimated by the sum of all negative eigenvalues.
\begin{eqnarray}
\sum_{i=1}^n \left[ T_i^{\alpha} - \frac{Z}{|x_i|}
  {\bf 1}_{\{|x_i|<r\}} \right]
- \frac{nZ}{r} & \ge & 
2 \, \mbox{Tr} \left[
T^{\alpha} - \frac{Z}{|x|}
  {\bf 1}_{\{|x|<r\}} \right]_-
- \frac{nZ}{r} \quad \mbox{for spin 1/2 fermions} \qquad \\
\sum_{i=1}^n \left[ T_i^{\alpha} - \frac{Z}{|x_i|}
  {\bf 1}_{\{|x_i|<r\}} \right]
- \frac{nZ}{r} & \ge &
n \, \mbox{Tr} \left[
T^{\alpha} - \frac{Z}{|x|}
  {\bf 1}_{\{|x|<r\}} \right]_-
- \frac{nZ}{r} \quad \mbox{without symmetry}. \qquad
\end{eqnarray}
We will use the combined Daubechies--Lieb--Yau
inequality\cite{Daub83b,Lieb88,Sol10} to finalize the proof of
Lemma 3.  Note again that our form differs from \cite{Sol10},
Theorem~16 (\textsl{Combined Daubechies--Lieb--Yau inequality}), in
setting $m=1$.\\

We state the combined DLY inequality in its application to the simpler
case of a single atom.  Assume a function $W \in
L^1_{loc}(\mathbb{R}^3)$ which satisfies
\begin{eqnarray}
W(x) \ge - \frac{\nu}{|x|} - C \nu \alpha^{-1} \qquad \mbox{when} 
\quad |x| < \alpha,
\end{eqnarray}
with $\alpha \nu \le 2/\pi$, $\alpha \ge 0$, and a constant
$C$. Then
\begin{eqnarray}
\mbox{Tr} \left[ T^{\alpha} + W(x) \right]_-
\ge && - C \nu^{5/2}\alpha^{1/2}
- C \int\limits_{\alpha <|x|} |W(x)_-|^{5/2} dx \nonumber \\
&& - C \alpha^3 \int\limits_{\alpha <|x|} |W(x)_-|^4 dx,
\end{eqnarray}
where, as above, $T^{\alpha} =
\sqrt{-\alpha^{-2}\, \Delta+\alpha^{-4}}-\alpha^{-1}$
and $T^{\alpha} = - \Delta/2$ when $\alpha = 0$.\\

With $\nu = Z$, we can evaluate
\begin{eqnarray}
\mbox{Tr} \left[ T^{\alpha} - \frac{Z}{|x|}
  {\bf 1}_{\{|x|<r\}} \right]_-
\ge && - C Z^{5/2} \alpha^{1/2} - C \int\limits_{|x|<r} 
\left(\frac{Z}{|x|}\right)^{5/2} dx \nonumber \\
&& - C \alpha^3 \int\limits_{\alpha <|x|<r} 
\left(\frac{Z}{|x|}\right)^{4} dx \nonumber \\
= && -C (Z\alpha)^{1/2} Z^2 - C Z^{5/2} r^{1/2} 
- C (Z \alpha)^2 Z^{2} \nonumber \\
\ge && -C Z^2 - C Z^{5/2} r^{1/2},
\end{eqnarray}
where we have used that $Z \alpha \le \frac{2}{\pi}$.\\

With the choice of $r$ of
\begin{eqnarray}
r & = Z^{-1} n^{2/3} & \quad \mbox{for fermions} \\
r & = Z^{-1} \phantom{n^{2/3}} & \quad \mbox{without symmetry,}
\end{eqnarray}
and using that $n \ge 1$, we complete the proof of Lemma~3. \\

\subsection{Proof of Theorem 4} 

Using Lemma~3 we can state
\begin{eqnarray}
( \Psi_n, \sum_{i=1}^n T_i^{\alpha} \Psi_n) 
- Z (\Psi_n, \sum_{i=1}^n \frac{1}{|x_i|} \Psi_n )
\ge - Z^2 \, \kappa(n).
\end{eqnarray}
We obtain an estimate on $( \Psi_n, \sum_{i=1}^n T_i^{\alpha} \Psi_n)$
through an assessment of the energy of a single atom
\begin{eqnarray}
0 \ge && E^{at}(n,Z;\alpha) = (\Psi_n, H^{at}(n,Z;\alpha) \Psi_n) \\
\ge && \varepsilon ( \Psi_n, \sum_{i=1}^n T_i^{\alpha} \Psi_n) 
+ (1 - \varepsilon) ( \Psi_n, \sum_{i=1}^n T_i^{\alpha} \Psi_n) 
- (\Psi_n, \sum_{i=1}^n \frac{Z}{|x_i|} \Psi_n ) \nonumber \\
\ge && \varepsilon ( \Psi_n, \sum_{i=1}^n T_i^{\alpha} \Psi_n) 
+ (1 - \varepsilon) ( \Psi_n, \sum_{i=1}^n \left( 
T_i^{\alpha} - \frac{Z}{(1-\varepsilon)|x_i|} \right)
\Psi_n) \nonumber \\
\ge && \varepsilon ( \Psi_n, \sum_{i=1}^n T_i^{\alpha} \Psi_n) 
- (1 - \varepsilon) \, \kappa(n) \frac{Z^2}{(1-\varepsilon)^{2}},
\end{eqnarray}
using again Lemma~3 in the last step, asserting that
$\frac{Z}{1-\varepsilon} \alpha \le \frac{2}{\pi}$ by the assumption
of Theorem~4. We can conclude that
\begin{eqnarray}
( \Psi_n, \sum_{i=1}^n T_i^{\alpha} \Psi_n) 
\le \frac{(1-\varepsilon)^{-1}}{\varepsilon} \kappa(n) \, Z^2,
\end{eqnarray}
and thereby complete the proof of Theorem~4.

\end{document}